\documentclass[prd,a4paper,superscriptaddress]{revtex4}
\usepackage{graphicx}

\begin{document}

\bibliographystyle{apsrev}

\newcommand {\R}{{\mathcal R}}
\newcommand{\al}{\alpha}

\title{New constraints on varying $\alpha$}

\author{G. Rocha}
\email[Electronic address: ]{graca@mrao.cam.ac.uk}
\affiliation{Astrophysics Group, Cavendish Laboratory,
Madingley Road, Cambridge CB3 0HE, United Kingdom}
\affiliation{Centro de Astrof\'{\i}sica da Universidade do Porto, R. das
Estrelas s/n, 4150-762 Porto, Portugal}
\author{R. Trotta}
\email[Electronic address: ]{trotta@amorgos.unige.ch}
\affiliation{D\'epartement de Physique Th\'eorique, Universit\'e de
Gen\`eve, 24 quai Ernest Ansermet, CH-1211 Gen\`eve 4, Switzerland}
\author{C.J.A.P. Martins}
\email[Electronic address: ]{C.J.A.P.Martins@damtp.cam.ac.uk}
\affiliation{Centro de Astrof\'{\i}sica da Universidade do Porto, R. das
Estrelas s/n, 4150-762 Porto, Portugal}
\affiliation{Department of Applied Mathematics and Theoretical Physics,
Centre for Mathematical Sciences,\\ University of Cambridge,
Wilberforce Road, Cambridge CB3 0WA, United Kingdom}
\affiliation{Institut d'Astrophysique de Paris, 98 bis Boulevard Arago,
75014 Paris, France}
\author{A. Melchiorri}
\email[Electronic address: ]{melch@astro.ox.ac.uk}
\affiliation{Department of Physics, Nuclear \& Astrophysics Laboratory,
University of Oxford, Keble Road, Oxford OX1 3RH, United Kingdom}
\author{P. P. Avelino}
\email[Electronic address: ]{pedro@astro.up.pt}
\affiliation{Centro de Astrof\'{\i}sica da Universidade do Porto, R. das
Estrelas s/n, 4150-762 Porto, Portugal}
\affiliation{Departamento de F\'{\i}sica da Faculdade de Ci\^encias
da Universidade do Porto, R. do Campo Alegre 687, 4169-007 Porto, Portugal}
\author{P.T.P. Viana}
\email[Electronic address: ]{viana@astro.up.pt}
\affiliation{Centro de Astrof\'{\i}sica da Universidade do Porto, R. das
Estrelas s/n, 4150-762 Porto, Portugal}
\affiliation{Departamento de Matem\'atica Aplicada da Faculdade de Ci\^encias
da Universidade do Porto, Rua do Campo Alegre 687, 4169-007 Porto, Portugal}


\begin{abstract}
We present a summary of recent constraints on the value of the
fine-structure constant at the epoch of decoupling from the recent
observations made by the Wilkinson Microwave Anisotropy Probe
(WMAP) satellite. Within the set of models considered, a variation of the value of $\alpha$ at decoupling with respect to the present-day value is now bounded to
be smaller than $2 \%$ ($6 \%$) at $95 \%$ confidence level.
We point out that the existence of an early reionization epoch as suggested 
by the
above measurements will, when more accurate cosmic microwave
background polarization data is available, lead to considerably
tighter constraints.
We find that the tightest possible constraint on $\alpha$ is
about $0.1 \%$ using CMB data alone---tighter constraints
will require further (non-CMB) priors.
\end{abstract}

\def\edth{\;\raise1.0pt\hbox{$'$}\hskip-6pt\partial\;}
\def\baredth{\;\overline{\raise1.0pt\hbox{$'$}\hskip-6pt
\partial}\;}
\def\bi#1{\hbox{\boldmath{$#1$}}}
\def\gsim{\raise2.90pt\hbox{$\scriptstyle
>$} \hspace{-6.4pt}
\lower.5pt\hbox{$\scriptscriptstyle
\sim$}\; }
\def\lsim{\raise2.90pt\hbox{$\scriptstyle
<$} \hspace{-6pt}\lower.5pt\hbox{$\scriptscriptstyle\sim$}\; }

\maketitle

\section{Introduction}

Cosmology and astrophysics provide a laboratory with extreme
conditions in which to test fundamental physics and search for new
paradigms. Currently preferred unification theories \cite{Polchinski}
predict the existence of additional
space-time dimensions, which have a number of possibly observable
consequences, including modifications in the gravitational laws on
very large (or very small) scales and space-time variations of the
fundamental constants of nature \cite{Uzan,Essay}. Recent evidence
of a time variation of fundamental constants
\cite{Webb,Jenam,Ivanchik} offers an important opportunity to test
such fundamental physics models. It should be noted that the issue
is not \textit{if} such theories predict such variations, but
\textit{at what level} they do so, and hence if there is
any hope of detecting them in the near future.

The most promising case is that of the fine-structure constant
$\alpha$, for which some evidence of time variation at redshifts
$z\sim2-3$ already exists \cite{Webb,Jenam}. Since one expects
$\alpha$ to be a non-decreasing function of time \cite{Damour,Barrow}, it
is particularly important to try to constrain it at earlier
epochs, where any variations relative to the present-day value
should be larger. The cosmic microwave background (CMB)
anisotropies provide such a probe, being mostly sensitive to the
epoch of decoupling, $z \sim 1100$.  \cite{Old,Avelino,Martins,Martinsw}

The reason the CMB is a good probe of
variations of the fine-structure constant is that these alter the
ionisation history of the universe \cite{steen,Kap,Old}.
The dominant effect is a change
in the redshift of recombination, due to a shift in the energy levels
(and, in particular, the binding energy) of Hydrogen. The Thomson
scattering cross-section is also changed for all particles, being
proportional to $\alpha^2$. A smaller effect
(which has so far been neglected) is expected to come from a change in
the Helium abundance.
As is well known, CMB fluctuations are typically described in terms
of spherical harmonics,
$T(\theta,\phi)=\sum_{\ell m}a_{\ell m}Y_{\ell m}(\theta,\phi)\ $
from whose coefficients one defines the angular power spectrum
$C_\ell=<|a_{\ell m}|^2>\ $. 
Increasing $\alpha$ increases the redshift
of last-scattering, which corresponds to a smaller sound horizon. Since the
position of the first Doppler peak ($\ell_{peak}$)
is inversely proportional to the sound horizon at last scattering, 
increasing $\alpha$ will produce a larger $\ell_{peak}$ \cite{Old}.
This larger redshift of last scattering also has the additional effect
of producing a larger early ISW effect, and hence a larger amplitude
of the first Doppler peak \cite{steen,Kap}. Finally, an increase
in $\alpha$ decreases the high-$\ell$ diffusion damping (which is
essentially due to the finite thickness of the last-scattering surface),
and thus increases the power on very small scales. 
These effects have been implemented in a modified CMBFAST algorithm which allows a varying $\alpha$ parameter \cite{Old,Avelino}. The changes were introduced in the subroutine RECFAST \cite{recfast} according to the extensive description given in \cite{steen,Kap}.

\section{Results and discussion}
In \cite{Martinsw} we presented a up-to-date constraints on the value of the
fine-structure constant at the epoch of decoupling from the recent
observations made by the Wilkinson Microwave Anisotropy Probe
(WMAP) satellite. In the framework of models considered, a
positive (negative) variation of the value of $\alpha$ at
decoupling with respect to the present-day value is now bounded to
be smaller than $2 \%$ ($6 \%$) at $95 \%$ C.L..

\begin{figure}
\includegraphics[width=3.5in]{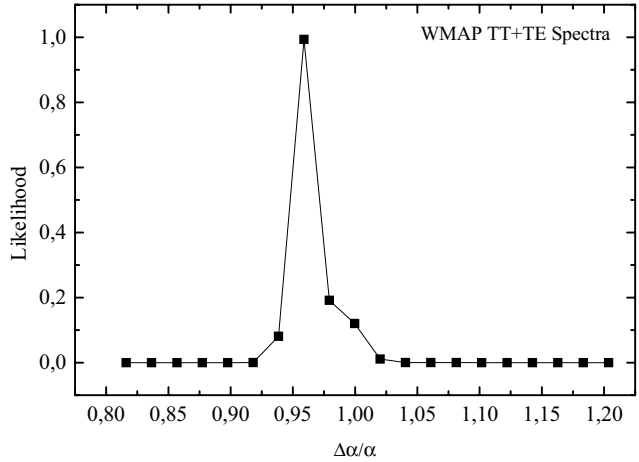}
\caption{\label{figalpha}
Likelihood distribution function for variations in the fine structure
constant obtained by an analysis of the WMAP data.}
\end{figure}

The likelihood
distribution function for $\alpha_{\text{dec}} / \alpha_0$,
obtained after marginalization over the remaining parameters, is
plotted in Figure \ref{figalpha}. We found, at $95 \%$ C.L. that $0.94 <
\alpha_{\text{dec}} / \alpha_0 < 1.01$, improving previous bounds,
(see \cite{Martins}) based on CMB and complementary datasets.


WMAP satellite data tightens the CMB constraints on the value of the
fine-structure constant at the epoch of decoupling. As in other previous works 
\cite{Old,Avelino,Martins}, the current data is
consistent with no variation,though the likelihood  is skewed
towards smaller values at the epoch of decoupling.

These previous works results were somewhat weakened by the existence of various important degeneracies in the data. This issue has been analysed by means of a Fisher Matrix Analysis (FMA) \cite{Martinsw,Rocha}.  
Following \cite{Rocha} we present the precision with which cosmological
parameters can be reconstructed using both CMB temperature and
E-polarization measurements. We consider the WMAP experiment, the planned
Planck satellite and an ideal experiment which would measure both
temperature and polarization to the cosmic variance limit (in the
following, 'CVL experiment`).
Cosmological models are characterized by the 8 dimensional
parameter set
\begin{equation}
{\bf \Theta} = (\Omega_b h^2, \Omega_m h^2, \Omega_\Lambda h^2,
\R, n_s, Q, \tau, \al),
\end{equation}
We assume  purely adiabatic initial conditions and we do not allow
for a tensor contribution. Our maximum likelihood model has
parameters $\omega_b = 0.0200$, $\omega_m = 0.1310$,
$\omega_\Lambda = 0.2957$ (and $h = 0.65$), $\R = 0.9815$, $n_s =
1.00$,$Q = 1.00$, $\tau=0.20$ and $\al/\al_0=1.00$.

The
experimental parameters used for the Planck analysis are in Table
\ref{exppar}, and we use the first 3 channels of the Planck High
Frequency Instrument (HFI) only. For the cosmic variance limited
(CVL) experiment, we set the experimental noise to zero, and we
use a total sky coverage $f_{\rm{sky}} = 1.00$. Although this is
never to be achieved in practice, the CVL experiment illustrates
the precision which can be obtained {\it in principle} from CMB
temperature and E-polarization measurements.


\begin{table}
\caption{\label{exppar} Experimental parameters for Planck
(nominal mission). Note that the sensitivities are here expressed
in $\mu$K.}
\begin{ruledtabular}
\begin{tabular}{|l|ccc|}
&  \multicolumn{3}{c|}{Planck HFI} \\
 \hline $\nu$ (GHz)  &
               $100$ &  $143$ & $217$  \\
$\theta_c$ (arcmin)&
    $10.7$ & $8.0$ & $5.5$ \\
$\sigma_{cT}$ ($\mu$K)  &
    $5.4$  & $6.0$  & $13.1$  \\
$\sigma_{cE}$ ($\mu$K)  &
    $n/a$  & $11.4$  & $26.7$  \\
$w^{-1}_c \cdot 10^{15}$ (K$^2$ ster) &
    $0.215$ & $0.158$ & $0.350$  \\
$\ell_c$ &
    $757$ & $1012$ & $1472$ \\\hline
$\ell_{\rm max}$ & \multicolumn{3}{c|}{$2000$} \\
$f_{\rm sky}$    & \multicolumn{3}{c|}{$0.80$} \\
\end{tabular}
\end{ruledtabular}
\end{table}

\begin{figure}
\includegraphics[width=3.5in,angle=90]{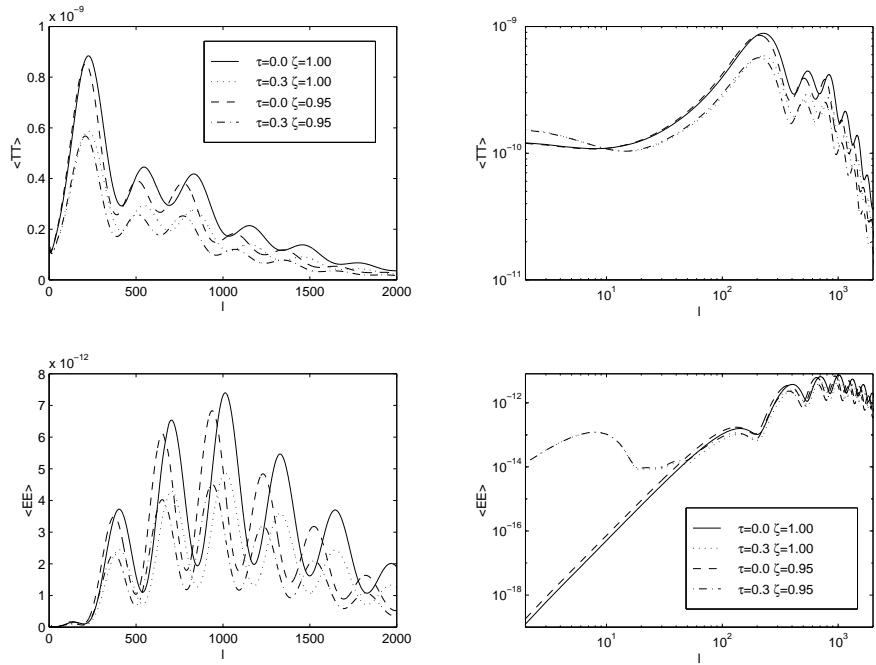}
\caption{\label{figcells} Contrasting the effects of varying
$\alpha$ and reionization on the CMB temperature and polarization.
Here $\zeta=\al_{dec}/\al_0$.}
\end{figure}

Fig.~\ref{figcells} illustrates the effect of $\alpha$ and $\tau$
on the CMB temperature and polarization power spectra---see
\cite{Rocha} for a more detailed discussion.


For the CMB temperature, reionization simply
changes the amplitude of the acoustic peaks, without affecting
their position and spacing (top left panel); a different value of
$\alpha$ at the last scattering, on the other hand, changes both
the amplitude and the position of the peaks (top right panel). The
outstanding effect of reionization is to introduce a bump in the
polarization spectrum at large angular scales (lower left panel).
This bump is produced well after decoupling at much lower
redshifts, when $\alpha$, if varying, is much closer to the
present day's value.  If the value of
$\alpha$ at low redshift is different from that at decoupling, the
peaks in the polarization power spectrum at small angular scales
will be shifted sideways, while the reionization bump on large
angular scales won't (lower right panel). It follows that by
measuring the separation between the normal peaks and the bump,
one can measure both $\alpha$ and $\tau$.
 Thus we expect that the existence of an early
reionization epoch will, when more accurate cosmic microwave
background polarization data is available, lead to considerably
tighter constraints on $\alpha$.

\begin{figure}
\includegraphics[width=3.5in]{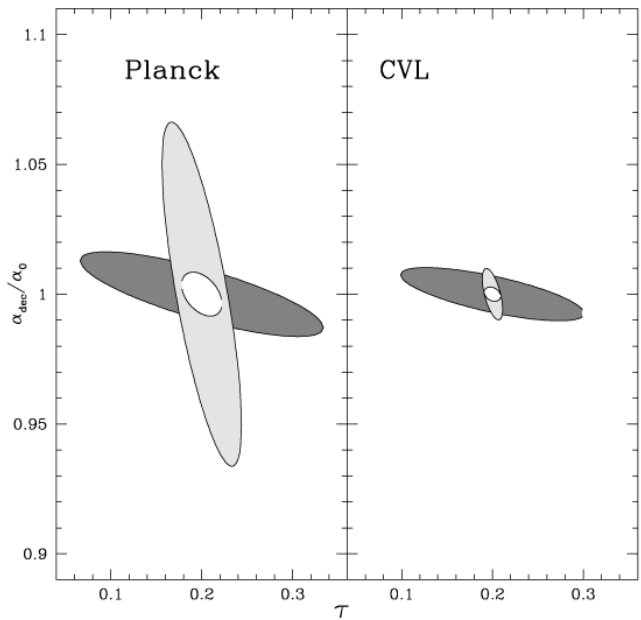}
\caption{\label{figlike} Ellipses containing $95.4\%$ ($2\sigma$)
of joint confidence in the $\alpha$ vs. $\tau$ plane (all other
parameters marginalized), for the Planck and cosmic variance
limited (CVL) experiments, using temperature alone (dark gray),
E-polarization alone (light gray), and both jointly (white).}
\end{figure}

Table \ref{fmaresults} and Fig.~\ref{figlike} summarize the
forecasts for the precision in determining $\tau$ and $\alpha$
(relative to the present day value) with Planck and the CVL
experiment. We consider the use of temperature information alone
(TT), E-polarization alone (EE) and both channels (EE+TT) jointly.
Note that one could use the temperature-polarization cross
correlation (ET) instead of the E-polarization, with the same
results. As it is apparent from Fig.~\ref{figlike}, TT and EE
suffer from degeneracies in different directions, because of the
reasons explained above. Thus combination of high-precision
temperature and polarization measurements can constrain in the
most effective ways both variations of $\alpha$ and $\tau$. Planck
will be essentially cosmic variance limited for temperature but
there will still be considerable room for improvement in
polarization (Table \ref{fmaresults}). This therefore argues for a
post-Planck polarization experiment, not least because
polarization is, in itself, better at determining cosmological
parameters than temperature. We conclude that Planck alone will be
able to constrain variations of $\alpha$ at the epoch of
decoupling within $0.34 \%$ ($1\sigma$, all other parameters
marginalized), which corresponds to approximately a factor 5
improvement on the current upper bound. On the other hand, the CMB
\textit{alone} can only constrain variations of $\alpha$ up to
${\cal O}(10^{-3})$ at $z \sim 1100$. Going beyond this limit
will require additional (non-CMB) priors on some of the other
cosmological parameters. This result is to be
contrasted with the variation measured in quasar absorption
systems by Ref.\cite{Webb}, $\delta \alpha / \alpha_0 = {\cal
O}(10^{-5})$ at $z \sim 2$. Nevertheless, there are models
where deviations from the present value could be detected
using the CMB.


\begin{table}
\caption{\label{fmaresults}Fisher matrix analysis results for a
model with varying $\alpha$ and reionization: expected $1\sigma$
errors for the Planck satellite and for the CVL experiment (see
the text for details). The column {\it marg.} gives the error with
all other parameters being marginalized over; in the column {\it
fixed} the other parameters are held fixed at their ML value; in
the column {\it joint} all parameters are being estimated
jointly.}
\begin{ruledtabular}
\begin{tabular}{|c| c c c|c c c|}
 &  \multicolumn{6}{c}{$1\sigma$ errors (\%)} \\\hline
             & \multicolumn{3}{c}{Planck HFI} & \multicolumn{3}{c}{CVL} \\
             & marg.  & fixed & joint  & marg.  & fixed & joint           \\\hline
     & \multicolumn{6}{c}{E-Polarization Only (EE)} \\\hline
$\alpha$       &2.66       &0.06       &7.62     &0.40        &$<0.01$        &1.14 \\
$\tau$         &8.81       &2.78       &25.19    &2.26 &1.52 &6.45
\\ \hline
     & \multicolumn{6}{c}{Temperature Only (TT)} \\\hline
$\alpha$       &0.66       &0.02       &1.88     &0.41        &0.01        &1.18  \\
$\tau$         &26.93      &8.28       &77.02    &20.32 &5.89
&58.11 \\ \hline
     & \multicolumn{6}{c}{Temperature + Polarization (TT+EE)} \\\hline
$\alpha$       &0.34        &0.02      &0.97     &0.11        &$<0.01$        &0.32 \\
$\tau$         &4.48        &2.65      &12.80    &1.80       &1.48        &5.15 \\
\end{tabular}
\end{ruledtabular}
\end{table}

\begin{figure}
\includegraphics[width=3.5in]{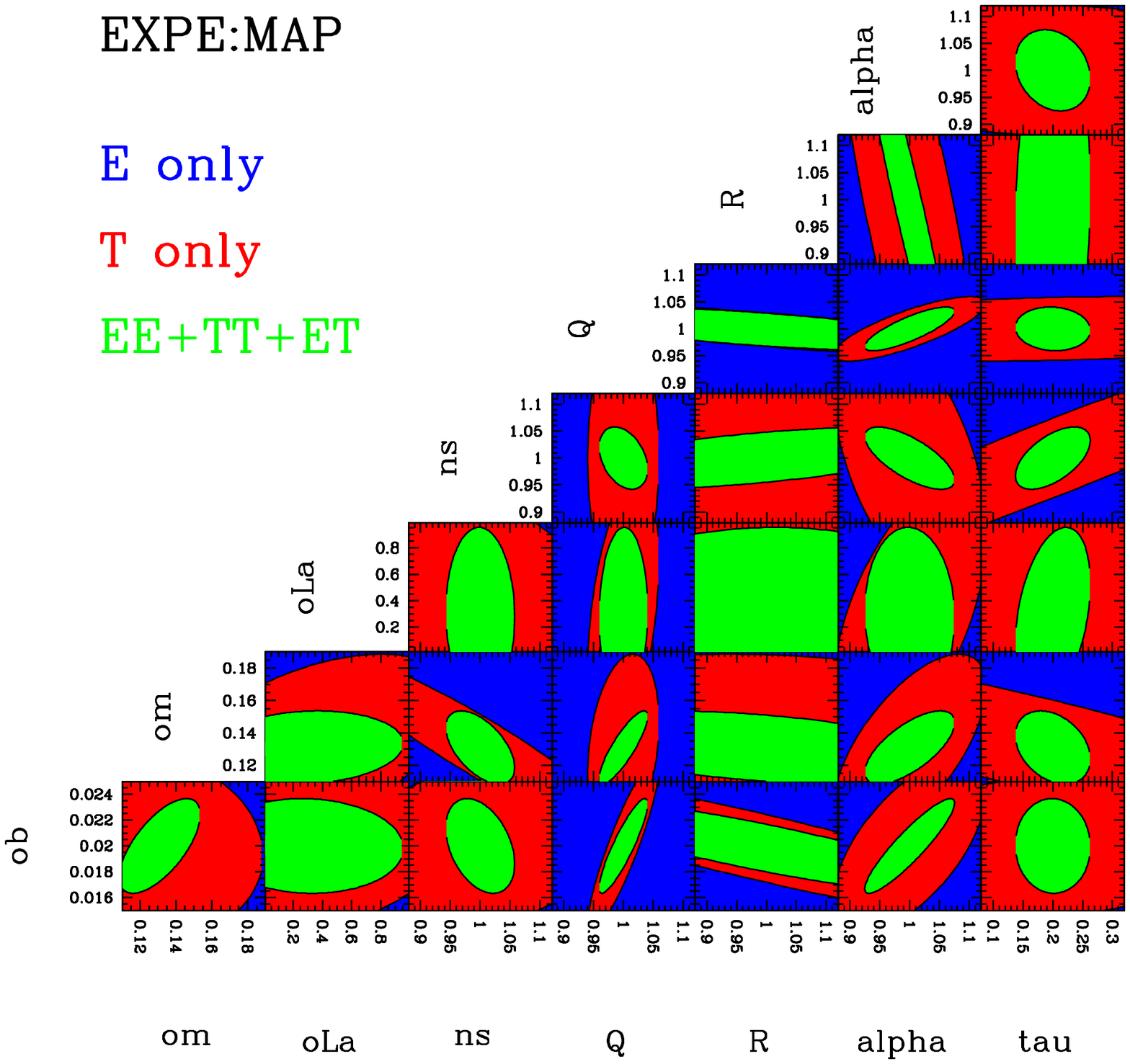}
\includegraphics[width=3.5in]{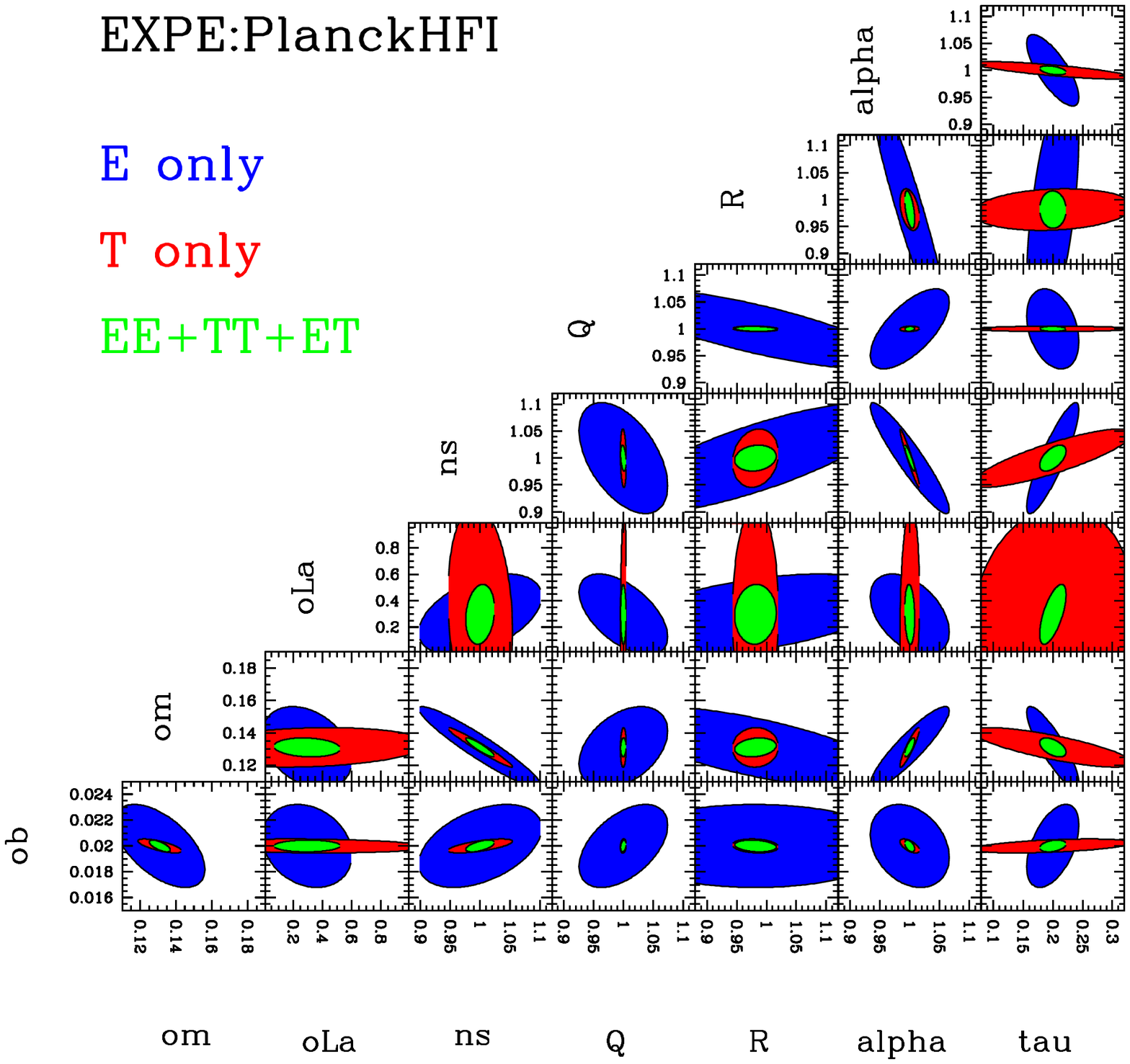}
\includegraphics[width=3.5in]{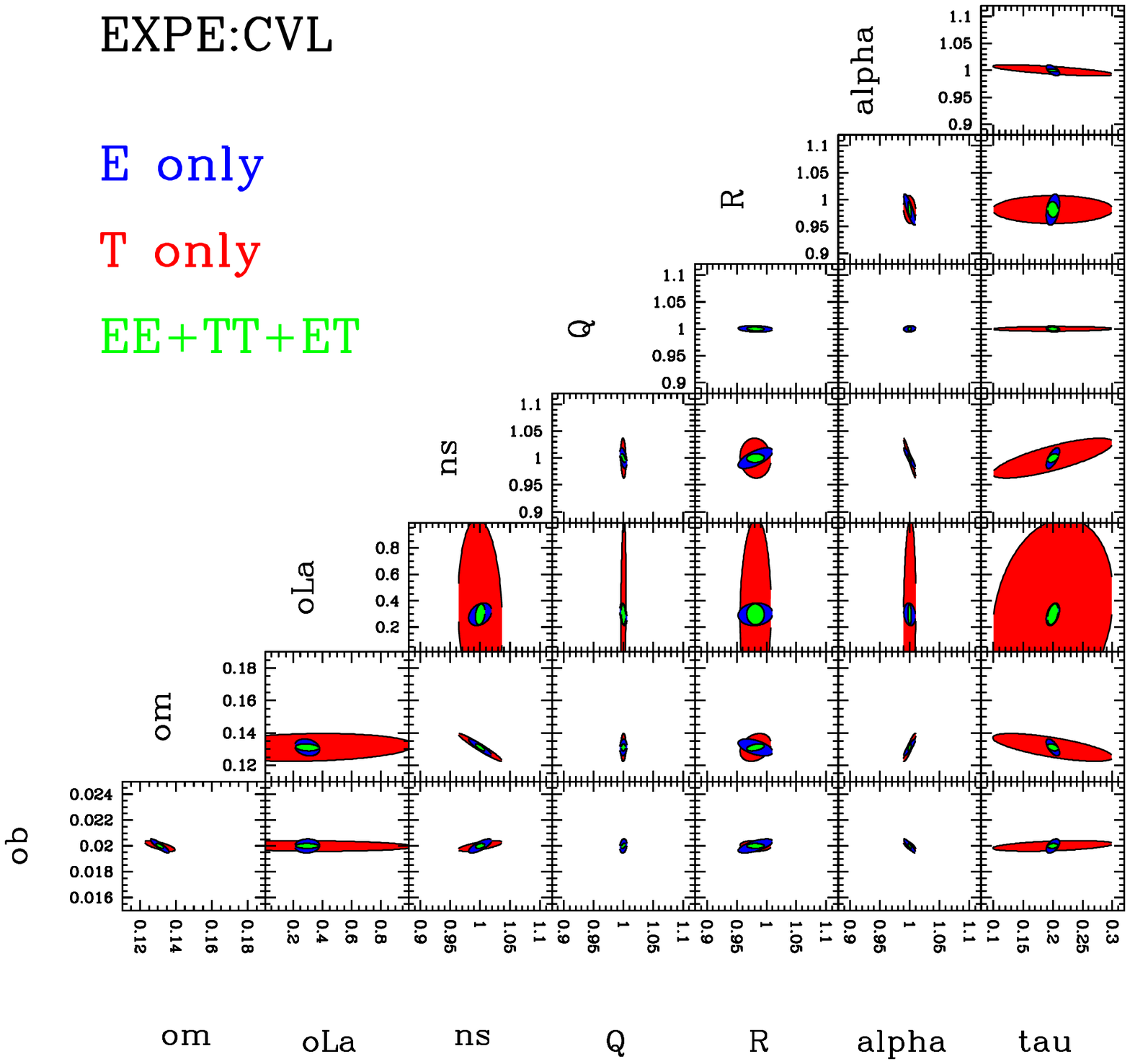}
\caption{\label{total} Ellipses containing $95\%$ ($2\sigma$) of
joint confidence (all other
parameters marginalized).}
\end{figure}

Finally Table \ref{fmarestotal} and  Fig.~\ref{total} summarize the FMA results for all parameters for WMAP, Planck and a CVL experiment see \cite{Rocha} for further details.

\section{Conclusions}
We presented up-to-date constraints on the value of the
fine-structure constant at the epoch of decoupling, using the WMAP satellite data.
Within the set of models considered, a variation of the value of $\alpha$ at
decoupling with respect to the present-day value is now bounded to
be smaller than $2 \%$ ($6 \%$) at $95 \%$ confidence level.
We have proposed a way of using the existence of an early reionization epoch as suggested by WMAP, to improve these constraints.
We have shown that CMB data alone will be able to constrain $\alpha$ 
up to the  $0.1 \%$ level. Tighter constraints than this will require
invoking further (non-CMB) priors.
These points are discussed in more detail in \cite{Rocha}.

\begin{table*}
\caption{\label{fmarestotal}Fisher matrix analysis results for a model with varying $\alpha$ and inclusion of reionization: expected $1\sigma$ errors for the MAP and Planck satellites as well as for a CVL experiment. The column {\it marg.} gives the error with all
other parameters being marginalized over; in the column {\it fixed} the other
parameters are held fixed at their ML value; in the column {\it joint} all
parameters are being estimated jointly.}
\begin{ruledtabular}
\begin{tabular}{|c|c c c| c c c|c c c|}
Quantity &  \multicolumn{9}{c}{$1\sigma$ errors (\%)} \\\hline  
             & \multicolumn{3}{c|}{MAP}           & \multicolumn{3}{c}{Planck HFI} & \multicolumn{3}{c}{CVL} \\ 
                        & marg. & fixed  & joint   & marg.  & fixed & joint  & marg.  & fixed & joint           \\\hline
	 & \multicolumn{9}{c}{Polarization} \\\hline
$\omega_b$     &281.91       &22.18      &806.27    &6.46       &1.11       &18.47    &1.09        &0.25        &3.12 \\
$\omega_m$     &446.89       &22.12      &1278.15   &7.75       &0.39       &22.17    &1.61        &0.03        &4.60 \\
$\omega_\Lambda$  &1248.94   &113.78     &3572.04   &41.61      &22.87      &119.01   &11.60       &9.99        &33.17 \\
$n_s$          &126.90       &5.31       &362.93    &4.14       &0.96       &11.85    &0.77        &0.08        &2.22 \\ 
$Q$            &200.97       &18.38      &574.78    &2.99       &0.51       &8.55     &0.24        &0.07        &0.68 \\ 
$\R$           &254.76       &20.44      &728.63    &9.56       &0.35       &27.33    &1.19        &0.03        &3.40 \\ 
$\alpha$       &111.52       &3.74       &318.96    &2.66       &0.06       &7.62     &0.40        &$<0.01$        &1.14 \\ 
$\tau$         &275.13       &9.64       &786.88    &8.81       &2.78       &25.19    &2.26        &1.52        &6.45 \\ \hline 
	 & \multicolumn{9}{c}{Temperature} \\\hline
$\omega_b$     &13.56        &1.35       &38.78     &1.09       &0.60       &3.12     &0.83        &0.38        &2.37  \\       
$\omega_m$     &17.73        &0.88       &50.71     &3.76       &0.13       &10.74    &2.64        &0.08        &7.55  \\        
$\omega_\Lambda$ &137.68     &96.36      &393.77    &111.61     &96.15      &319.21   &98.97       &86.00       &283.05 \\       
$n_s$          &10.10        &0.53       &28.88     &2.18       &0.13       &6.24     &1.49        &0.07        &4.26  \\        
$Q$            &2.41         &0.36       &6.89      &0.20       &0.11       &0.57     &0.18        &0.07        &0.50  \\       
$\R$           &23.86        &0.78       &68.25     &1.58       &0.12       &4.53     &1.06        &0.07        &3.04  \\       
$\alpha$       &5.16         &0.13       &14.76     &0.66       &0.02       &1.88     &0.41        &0.01        &1.18  \\     
$\tau$         &111.97       &13.26      &320.24    &26.93      &8.28       &77.02    &20.32       &5.89        &58.11 \\ \hline 
	 & \multicolumn{9}{c}{Temperature and Polarization} \\\hline
$\omega_b$     &7.37         &1.34       &21.07     &0.91        &0.53      &2.61     &0.38        &0.21        &1.09  \\
$\omega_m$     &6.94         &0.88       &19.85     &1.81        &0.12      &5.17     &0.67        &0.03        &1.91 \\
$\omega_\Lambda$  &89.69     &72.75      &256.51    &30.89       &22.04     &88.36    &10.79       &9.85        &30.85\\
$n_s$          &2.32         &0.52       &6.65      &0.97        &0.13      &2.77     &0.33        &0.05        &0.93 \\
$Q$            &1.63         &0.36       &4.67      &0.19        &0.10      &0.54     &0.14        &0.05        &0.41\\
$\R$           &14.22        &0.78       &40.68     &1.43        &0.11      &4.08     &0.60        &0.03        &1.72 \\
$\alpha$       &3.03         &0.13       &8.68      &0.34        &0.02      &0.97     &0.11        &$<0.01$        &0.32 \\
$\tau$         &12.67        &7.90       &36.23     &4.48        &2.65      &12.80    &1.80        &1.48        &5.15 \\
\end{tabular}
\end{ruledtabular}
\end{table*}

\bibliography{cmbnet-03}

\end{document}